\documentclass[letter,twocolumn]{jpsj2}
\title{Numerical Study of Current-Induced Domain-Wall Dynamics: 
Crossover from Spin Transfer to Momentum Transfer}

\author{Daisuke \textsc{Matsubayashi}$^{1}$\thanks{E-mail: dmatsuba@hosi.phys.s.u-tokyo.ac.jp.}, Masafumi  \textsc{Udagawa}$^{2}$, and Masao \textsc{Ogata}$^{1}$}

\inst{$^{1}$Department of Physics, University of Tokyo, Hongo, Bunkyo-Au, 
Tokyo, 113-0033 \\
$^{2}$Department of Applied Physics, University of Tokyo, Hongo, Bunkyo-Au, 
Tokyo, 113-8656}
\abst{We study current-induced dynamics of a magnetic domain
wall by solving a time-dependent Schr\"{o}dinger equation combined with 
Landau-Lifshitz-Gilbert equation in a one-dimensional 
electron system coupled to localized spins. Two types of
domain-wall motions are observed 
depending on the hard-axis anisotropy, $K_{\perp}$, of the
localized spin system. 
For small values of $K_{\perp}$, the magnetic domain wall shows 
a streaming motion driven by spin transfer. 
In contrast, for large values of $K_{\perp}$, 
a stick-slip motion driven by momentum transfer is obtained. 
We clarify the origin of these characters of domain-wall motions 
in terms of the dynamics
of one-particle energy levels and distribution functions.
}

\kword{domain wall, spin transfer, momentum transfer, stick-slip motion, 
time-dependent Schr\"{o}dinger equation}

\begin{document}
\maketitle
\begin{figure}[b]
\begin{center}
\includegraphics[width=5cm]{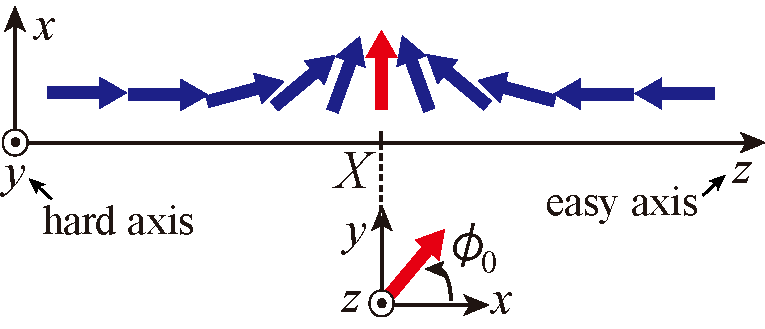}
\end{center}
\caption{(Color online) 
The schematic figure of a DW formed by localized spins: X and $\phi_0$ are
the displacement and the rotating angle of DW, respectively.}
\label{DW}
\end{figure}

A magnetic domain wall (DW) is a twisted structure of localized spins, which
separates the magnetic domains with different polarization in a
ferromagnet. Dynamics of topological defects, such as DWs and vortices, 
has been one of the central issues in condensed matter physics. 
The lowest excitations of polyacetylene are described as solitons, 
which determine the conductive properties
of the system\cite{Heeger}. The critical current of type-II superconductors is
sensitive to the pinning of vortices\cite{Tinkham}. 
Recently, current-induced dynamics of magnetic DWs attracts
considerable attention for its importance as a non-equilibrium many-body
problem and its potential for technical application 
\cite{tatara_rev, Berger2, Slonczewski}.
In particular, promising proposals have been made for the application to
non-volatile memory\cite{parkin}.
Nano-scale fabrication of the memory device may be
possible, by controlling the DW motion by an electric current.
For simplicity, 
let us consider a situation where there is a single DW separating
the $+z$-spin and the $-z$-spin regions (Fig. \ref{DW}), 
which is coupled to electrons ferromagnetically. 
Then suppose that an electron wave packet with $+z$-spin is 
injected from the left-hand side. 
In this case, one can imagine two mechanisms 
for the dynamics of the DW\cite{berger,tatara1}. 
(i)The spin of electron wave packet staying in the $+z$-spin region is
aligned in the $+z$ direction due to the ferromagnetic coupling to the
localized spins. If this electron is carried adiabatically to the
$-z$-spin region, the direction of electron spin gradually changes
and ends up in $-z$.
Due to the conservation of angular momentum, 
the total $z$-axis spin of localized spins 
increases by $\hbar$, which leads to the
enlargement (shrinking) of the $+z$ ($-z$)-spin region, 
resulting in a shift of the DW. 
This adiabatic process is called ``spin transfer." 
(ii)On the other hand, if the electron is transferred to the
$-z$-spin region with a finite velocity, 
the electron is reflected by the DW with a finite probability. 
Through this reflection, the DW acquires
a finite momentum as a result of recoil effect. 
This process is called ``momentum transfer,'' and originates from the
non-adiabatic motion of the conduction electrons, 
in contrast to the ``spin transfer."

The character of DW motion becomes qualitatively different depending on
which mechanism is dominant.
So far, extensive theoretical studies 
\cite{tatara1,barnes,thiaville,li,ohe,xiao,nguyen}
have shown that the ``spin transfer" is the dominant mechanism
for the DW motion in the experimental situations. 
Actually, many experimental results have been explained 
in this mechanism. 
\cite{yamaguchi,yamanouchi1,klaui,thomas,yamanouchi2}.
On the other hand, the relevance of ``momentum transfer'' has been
also reported in some experiments \cite{saitoh,hayashi}. 
The dynamics of thin DW also draws attention\cite{Feigenson}, in which
more careful treatment is needed for the effect of non-adiabaticity of 
conduction electrons, i.e., the effect of ``momentum transfer.''

Most of the theoretical studies 
have implicitly assumed that the conduction
electrons obey a static distribution function slightly off the
equilibrium state.
Under such assumption, the non-adiabaticity of
conduction electrons is not correctly taken into account.
In particular, it is expected that the effect of ``momentum transfer''
is underestimated. 
Therefore, in this paper, we will treat the dynamics of
conduction electrons and a DW on an equal footing, 
and discuss how the DW motion is affected 
by the strong non-adiabaticity of the conduction electron system.

Our model consists of a one-dimensional electron system 
under an electric field, 
which is coupled to a classical localized spin system,  
\begin{equation}
\begin{split}
\mathcal{H}(t) &= -t_\mathrm{c} \sum_{i,\sigma}
[\mathrm{e}^{\mathrm{i}\Phi(t)} c_{i+1\sigma}^{\dagger}
c_{i\sigma}+\mathrm{h.c.}]\\
&\quad-J_{\mathrm{sd}}\sum_{i,\sigma,\sigma'}
\boldsymbol{S}_i(t)\cdot c_{i\sigma}^{\dagger}
\boldsymbol{\sigma}_{\sigma\sigma'}c_{i\sigma'}.
\end{split}
\label{Hamiltonian}
\end{equation}
Here, $c_{i\sigma}$($c_{i\sigma}^{\dagger}$) is an annihilation
(creation) operator of the conduction electron at the $i$th site 
and spin $\sigma$
under the periodic boundary condition, 
$c_{i+N\sigma}=c_{i\sigma}$. 
$\boldsymbol{S}_i(t)$
denotes classical localized spins at the sites $i$, 
which constitute DWs with an easy z-axis and a hard y-axis. 
The electric field is applied via a time-dependent AB flux, $\Phi(t)$. 
In order to apply a direct electric field, we set $\Phi(t)= \frac{eV}{\hbar N}t$, 
where $V$ is the voltage drop in the circuit. 

We study the dynamics of state vector of conduction
electrons, $\left|\Psi(t)\right\rangle$, and the localized spins,
$\boldsymbol{S}_i(t)$, simultaneously.
$\left|\Psi(t)\right\rangle$ is determined by the time-dependent
Schr\"{o}dinger equation, 
$\mathrm{i}\hbar \frac{d}{dt}\left|\Psi(t)\right\rangle = 
\mathcal{H}(t)\left|\Psi(t)\right\rangle$. 
At the same time, we determine $\boldsymbol{S}_i(t)$ 
by solving the semiclassical Landau-Lifshitz-Gilbert (LLG) equations, 
\begin{equation}
\dot{\boldsymbol{S}_i} = 
\boldsymbol{H}_{\mathrm{eff}i}\times \boldsymbol{S}_i 
-\alpha (\boldsymbol{S}_i/S)\times \dot{\boldsymbol{S}_i}. 
\label{LLG}
\end{equation} 
The first term of eq. (\ref{LLG}) represents the torque which leads to   
the precessional motion of the localized spin 
around the effective magnetic field, 
$\boldsymbol{H}_{\mathrm{eff}i}$. The second term corresponds to
the Gilbert damping, which is essential to the relaxation of
$\boldsymbol{S}_i$ parallel to $-\boldsymbol{H}_{\mathrm{eff}i}$. Here, 
\begin{equation}
\begin{split}
\boldsymbol{H}_{\mathrm{eff}i} &= 
-J(\boldsymbol{S}_{i-1}+\boldsymbol{S}_{i+1})
-KS_{iz}\boldsymbol{e}_z\\
&\quad+K_{\perp}S_{iy}\boldsymbol{e}_y
-J_\mathrm{sd}\left\langle \boldsymbol{\sigma_i} \right\rangle, 
\label{H_eff}
\end{split}
\end{equation}
with the exchange coupling between localized spins $J$, 
the easy-axis anisotropy $K$, 
the hard-axis anisotropy $K_\perp$, and 
the electron spin 
$\left\langle \boldsymbol{\sigma}_i(t) \right\rangle$
$=\left\langle \Psi(t)\right| 
\sum_{\sigma,\sigma'}c_{i\sigma}^{\dagger}
\boldsymbol{\sigma}_{\sigma\sigma'}c_{i\sigma'}
\left|\Psi(t)\right\rangle$ at the $i$th site.

In order to solve the time-dependent Schr\"{o}dinger equation
numerically, we use the Crank-Nicholson's method \cite{crank},  
$\left|\Psi (t+\varDelta t)\right\rangle$
$= \exp [-\frac{\mathrm{i}}{\hbar}\int_t^{t+\varDelta t}\mathcal{H}(t')dt']
\left|\Psi(t)\right\rangle$
$\simeq \frac{1-(\mathrm{i}/\hbar)
\mathcal{H}(t+\varDelta t/2)\varDelta t/2}
{1+(\mathrm{i}/\hbar)
\mathcal{H}(t+\varDelta t/2)\varDelta t/2}
\left|\Psi(t)\right\rangle$, 
which ensures the norm conservation of the state vector. 
For the LLG equations, we use the implicit fourth-order Runge-Kutta
method. 

We set the initial configuration ($t=0$) of the localized spins to be
the static solution of the LLG equations, eq.\ (\ref{LLG}), with two DWs:
$\boldsymbol{S}_i(0)$
$=S\{1/\cosh[\frac{i_{c1}(0)-i}{W}], 0, 
\tanh[\frac{i_{c1}(0)-i}{W}]\}$ 
for the sites $1 \leq i$
$\leq N/2$, and 
$\boldsymbol{S}_i(0)=$
$[-S_{i-N/2 x}(0), S_{i-N/2 y}(0), -S_{i-N/2 z}(0)]$
for $N/2+1 \leq i \leq N$.
Here, $i_{c1}(0)=(N+2)/4$ 
is the initial coordinate of the first DW center, 
and $W$ is the width of the DWs.  
Note that an even number of DWs are required 
for smooth connection at the boundary 
because of the periodic boundary condition. 
Since the relation $\boldsymbol{S}_{i+N/2}(t)=[-S_{i x}(t),
S_{i y}(t), -S_{i z}(t)]$ holds
within the simulation time, 
we only focus on the dynamics of the first DW. We assume the initial state vector,
$\left|\Psi (0)\right\rangle$, to be the ground state of Hamiltonian,
eq. (\ref{Hamiltonian}), with the initial configuration,
$\boldsymbol{S}_i(0)$.

\begin{figure}[ht]
\begin{center}
\includegraphics[width=5cm]{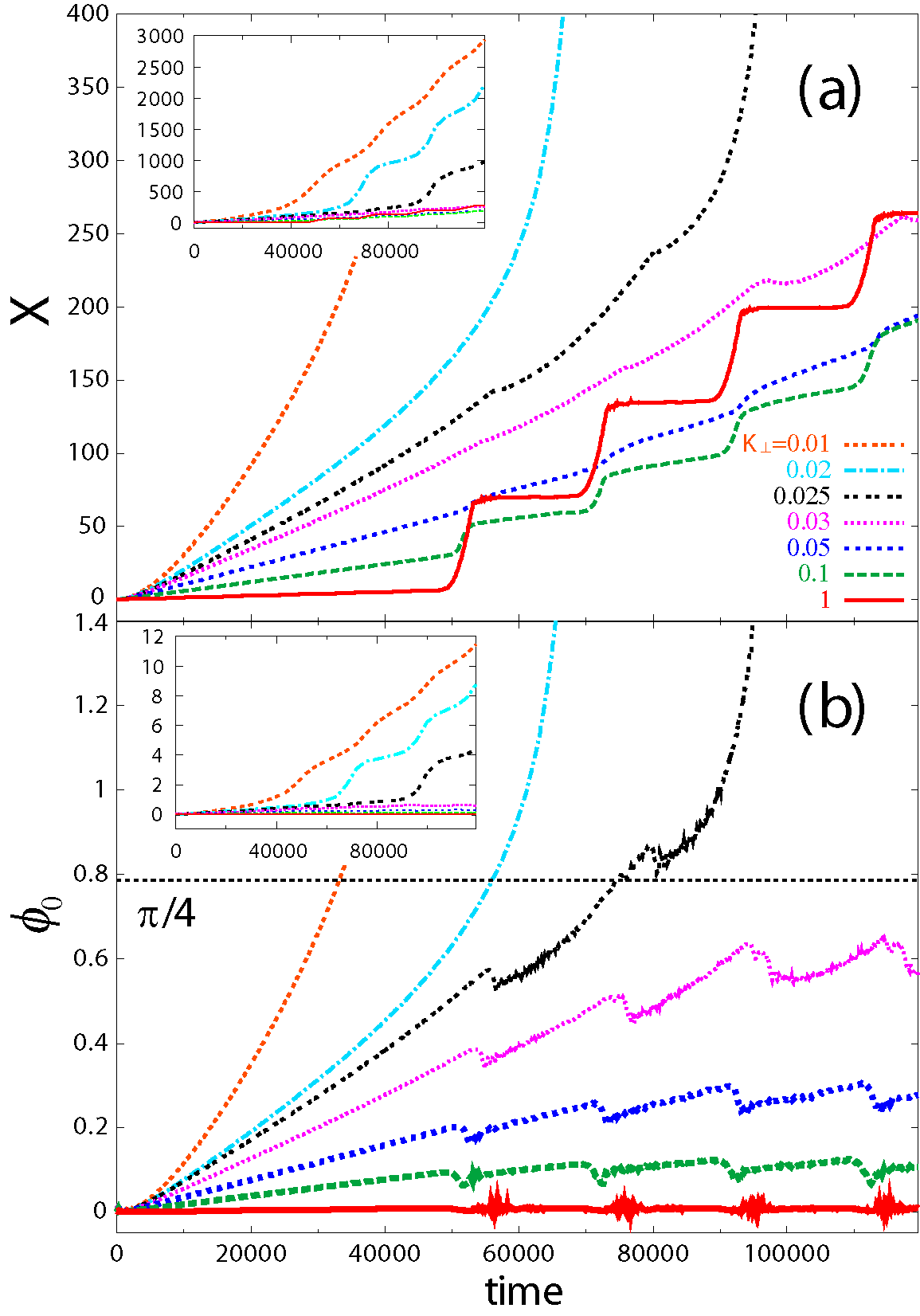}
\end{center}
\caption{(Color online) 
The time dependence of (a)the displacement $X$, and (b)the rotating angle
$\phi_0$ of the DW 
for several values of the hard-axis anisotropy $K_\perp$. 
The horizontal line of (b) corresponds to $\phi_0=\pi/4$. 
The scale of the longitudinal axis 
is expanded in the insets of (a) and (b).
} 
\label{X-p}
\end{figure}

\begin{figure}[ht]
\begin{center}
\includegraphics[width=5cm]{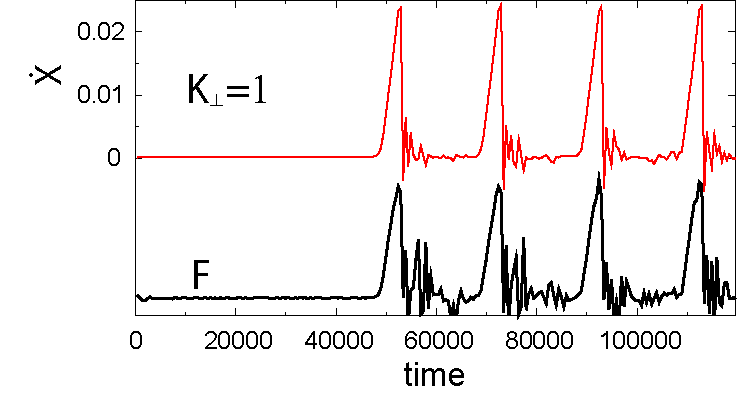}
\end{center}
\caption{(Color online)
The time dependence of the velocity, $\dot{X}$,  
and the force exerted by the electrons, $F$, for $K_\perp = 1$. 
The amplitude of $F$ is plotted in arbitrary unit.
}
\label{t-v-F}
\end{figure}

In order to characterize the DW motion, we define the displacement of DW
center, $X(t)$, and the rotating angle of 
the localized spins in the DW center around z-axis, $\phi_0(t)$,  
as shown in Fig. \ref{DW}. 
We will calculate the z-component of spin torque, 
$\tau_z(t) = J_\mathrm{sd}\sum_{i\in \mathrm{1st DW}}
[\boldsymbol{S}_i(t)\times\left\langle \boldsymbol{\sigma_i}(t)
\right\rangle]_z$. Since the first term in the right-hand side of 
eq. (\ref{LLG}) involves the
term proportional to $\tau_z(t)$, $\dot{S}_{iz}$ increases with $\tau_z(t)$.  
This means that $\tau_z(t)$ transfers z-axis angular
momentum from conduction electrons to localized spins, and is essential
to the ``spin transfer'' mechanism.
We also calculate 
the force, 
$F(t) =-J_\mathrm{sd}\sum_{i\in \mathrm{1st DW}} \nabla\boldsymbol{S}_i(t)\cdot
\left\langle \boldsymbol{\sigma_i}(t) \right\rangle$, 
which is the time variation of the momentum passed 
from the electron system to the localized spin system.  
Note that only if the DW has mass, $F(t)$ 
can be identified as a `real' force exerted by 
the conduction electrons on the DW and    
is relevant to the ``momentum transfer'' mechanism.
As shown below and derived in Ref. [\citen{tatara1}], 
both $\tau_z(t)$ and $F(t)$ determine the dynamics of DW.
In the following, we assume that 
$t_\mathrm{c}$ and $\hbar/t_\mathrm{c}$ are 
the units of energy and time, respectively, 
and that $\hbar=e=S=1$.

Here, we will show our results. The time evolution of $X(t)$ and 
$\phi_0(t)$ are shown in Figs. \ref{X-p}(a) and \ref{X-p}(b), respectively, 
for several values of $K_\perp$.   
We set $N=60$, $W=5$, $J=2.59$, $K=0.1$, $\alpha=0.02$, $J_\mathrm{sd}=0.3$, 
$V=\pi\times 10^{-4}$, and the electron density $n=12/N=0.2$. 
As shown in Figs. \ref{X-p}(a) and \ref{X-p}(b), 
the DW motion qualitatively changes,
depending on the value of $K_\perp$.

Firstly, for small values of $K_\perp$ ($K_\perp =$ 0.01, 0.02, 0.025), 
$X(t)$ and $\phi_0(t)$ increase smoothly with $t$ as shown in the insets 
of Fig. \ref{X-p}. 
The constant increase of $\phi_0(t)$ is caused
by the spin torque $\tau_z(t)$ with the help of Gilbert
damping\cite{torqueeffect}. 
This streaming motion of the DW accompanied by the continuous increase 
of $\phi_0(t)$ means dominance of 
the ``spin transfer" mechanism\cite{tatara1}.

Secondly, for large values of $K_\perp$ ($K_\perp = $ 0.03, 0.05, 0.1, 1), 
step-like increase of $X(t)$ with $t$ is observed in Fig. \ref{X-p}(a). 
In this case,  
the rotating angle $\phi_0(t)$ is saturated at a finite value 
in a longer time scale
in contrast to the case of smaller $K_\perp$. 
We plot the velocity $\dot{X}(t)$ and the force $F(t)$
for $K_\perp=1$ in Fig. \ref{t-v-F}. 
Remarkably, $\dot{X}(t)$ and $F(t)$ show peaks at
the same time.
This perfect correspondence clearly shows that the DW motion 
is driven by the force given by conduction electrons, i.e., 
the ``momentum transfer" is the dominant driving mechanism in this region. 
We will later discuss the microscopic origin of the periodic peak
structure of $F(t)$.

Here, we discuss the origin of crossover in driving mechanism
between ``spin transfer'' and ``momentum transfer.''
The crossover results from the competition between the spin torque
$\tau_z(t)$ and the hard-axis anisotropy $K_\perp$.
$K_\perp$ tends to fix $\phi_0(t)\simeq 0$, while $\tau_z(t)$ increases
$\phi_0(t)$ as we mentioned in the footnote [\citen{torqueeffect}].
The origin of the crossover behavior can be interpreted using 
the effective equations of motion for $X(t)$ and $\phi_0(t)$,\cite{tatara1}   
\begin{align}
&2(-\dot{\phi}_0+\alpha\dot{X}/W)=F,
\label{eom1}\\ 
&2(\dot{X}+\alpha W \dot{\phi}_0)=\tau_z-K_\perp W\sin 2\phi_0, 
\label{eom2}
\end{align} 
where irrelevant coefficients are omitted. 
For $\tau_z(t) >  K_\perp W$, 
since the right-hand side of eq. (\ref{eom2}) is always finite, 
constant supply of the spin torque $\tau_z(t)\gg \alpha W F(t)$ 
keeps both $X(t)$ and $\phi_0(t)\sim\alpha X(t)/W$ increasing.   
This is why the DW motion for smaller $K_\perp$
is driven by the ``spin transfer'' 
(the insets in Fig. \ref{X-p}).   
In contrast, for $\tau_z(t) <  K_\perp W$, 
the effect of the spin torque is 
canceled by the contribution of large $K_\perp$, 
and then $\phi_0(t)$ is saturated at the angle 
$\phi_\mathrm{s}=\frac{1}{2}\sin^{-1}\bigl(\frac{\tau_z}{K_{\perp}W}\bigr)
 (< \pi/4)$ in a longer time scale 
($K_\perp \ge 0.03$ in Fig. \ref{X-p}(b)). 
This is the so-called intrinsic pinning effect 
for the ``spin transfer'' mechanism.  
In the case of $\tau_z(t) \ll  K_\perp W$, 
since $\phi_0(t)\ll \pi/4$ and mass of the DW is well-defined  
as $M_\mathrm{DW} = \frac{2(1+\alpha^2)}{K_\perp W}$,
the ``momentum transfer'' from the conduction electrons 
becomes relevant to the DW motion. 

With $\tau_z(t) \ll  K_\perp W$ in mind, 
substituting $\phi_0(t)=\phi_\mathrm{s}+\delta\phi(t)$ 
into eqs. (\ref{eom1}) and (\ref{eom2}), 
we obtain the equations, $2(-\delta\dot{\phi}+\alpha\dot{X}/W)=F$, 
and $\dot{X}+\alpha W\delta\dot{\phi}=-K_\perp W \delta\phi$. 
The solutions are estimated as 
$\delta \dot{\phi}(t)=-\frac{1-c}{2(1+\alpha^2)}F(t)$ and 
$\dot{X}(t) = \frac{W(c+\alpha^2)}{2\alpha(1+\alpha^2)}F(t)$, 
including a numerical factor, $0<c<1$.
\cite{factor_c}  
In our calculation for $K_\perp=1$, $c \sim 0.95$ is obtained by
numerically averaging the ratio of $\dot{X}(t)$ and $F(t)$ over the
interval:\ $4\times 10^{4}<t<1.2\times 10^{5}$. 
Since $\alpha$ is small in general, 
the contribution of the force to the DW motion is remarkably 
enhanced, being inversely proportional to $\alpha$.\cite{tatara2} 
The fluctuation of $\phi_0(t)$ around the saturation angle $\phi_\mathrm{s}$ 
is crucial for the enhancement of the DW motion: $\dot{X}\propto \alpha^{-1}$.

Our calculation shows the absence of the perfect intrinsic pinning 
in a large $K_{\perp}$ region. 
While $\phi_0(t)$ is saturated for $K_\perp \ge 0.03$, 
the DW still shows a step-like motion. 
Absence of the intrinsic pinning has also been reported in the numerical 
study of ref. [\citen{ohe}]. However, they attribute the origin of DW motion
to the non-local deformation of localized spins, 
which is different from our results. 
In our calculation, deformation of localized spins is negligible. 

\begin{figure}[ht]
\begin{center}
\includegraphics[width=7cm]{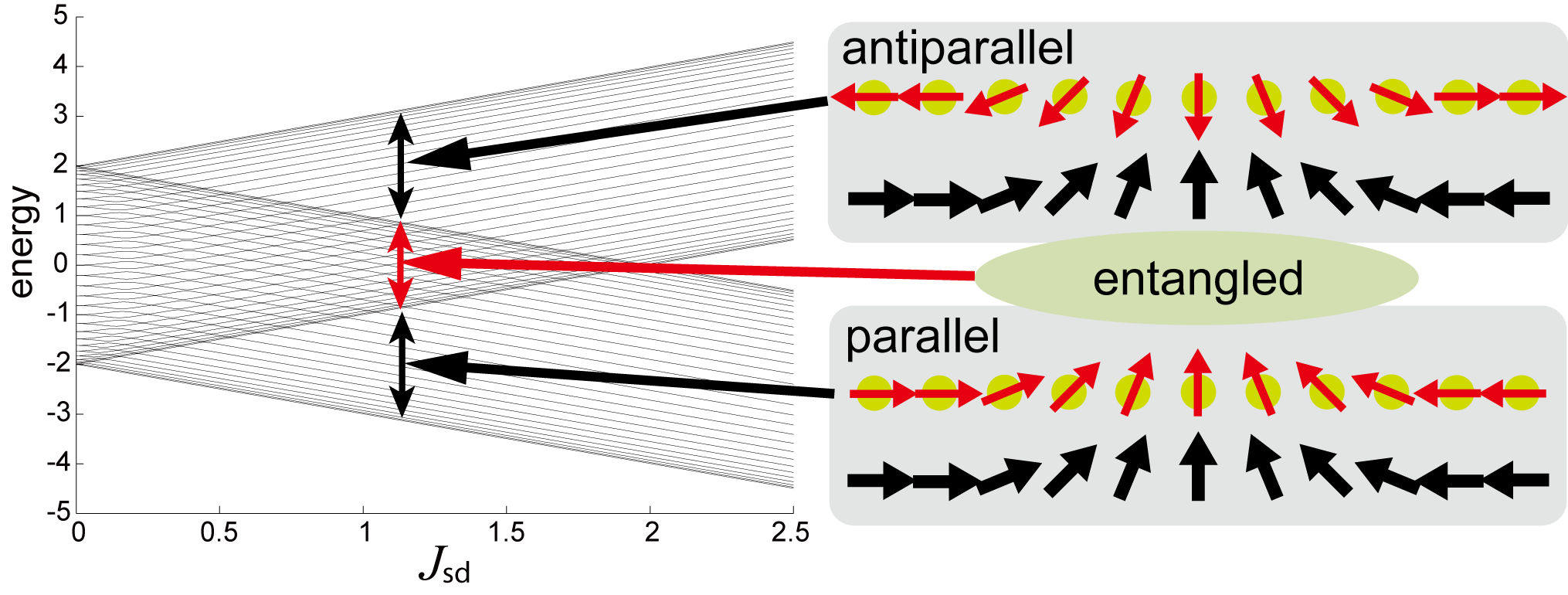}
\end{center}
\caption{(Color online) 
One-particle energy levels $\{E_i(0)\}$ for various $J_\mathrm{sd}$ values. 
The energy levels split into two bands. 
For $J_\mathrm{sd}\lesssim \Omega/2$, the two bands overlap at 
$|E_i(0)| <-J_\mathrm{sd}+\Omega/2$ 
where spin and momentum degrees of freedom are entangled, and a 
complicated energy level structure is formed.
}
\label{level_st}
\end{figure}

\begin{figure}[ht]
\begin{center}
\includegraphics[width=4.5cm]{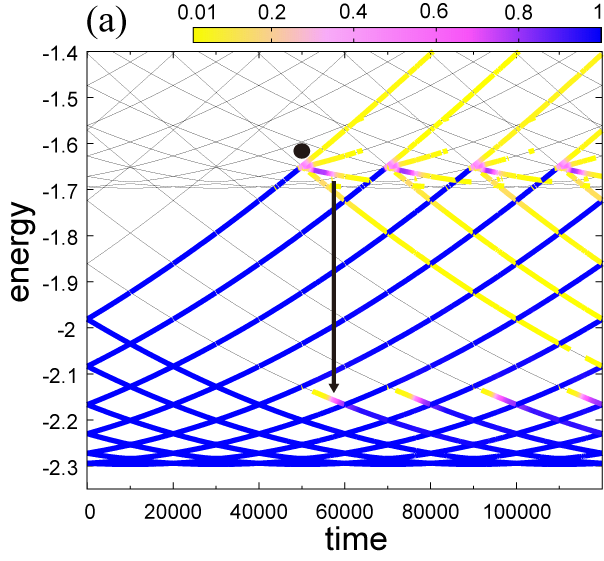}
\includegraphics[width=3.9cm]{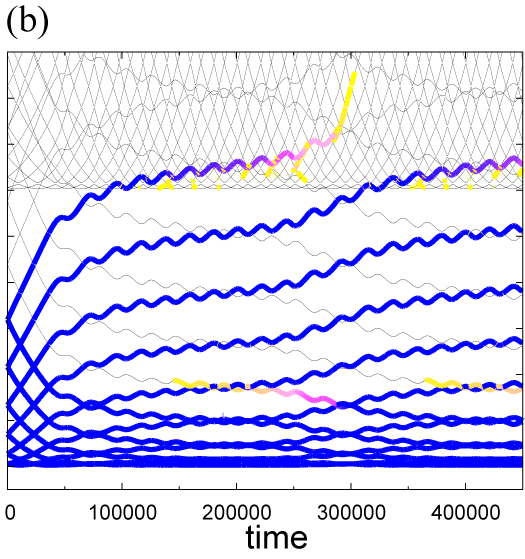}
\end{center}
\caption{(Color online) 
The time dependence of 
the one-particle energy levels $\{E_i(t)\}$ (thin lines) and 
the one-particle distribution functions 
$\{|\left\langle\xi_i(t)|\Psi(t)\right\rangle|^2\}$ (color plots) 
of the electron system for $K_\perp=$ (a) $1$ and (b) $0.01$. 
Note that the time scale in (b) is much larger than that in (a).  
The color of the thick lines represents 
the occupation probability of each level. 
The dot and the arrow in (a) denote an elastic backscattering 
and an inelastic scattering,\cite{inelastic} respectively. 
}
\label{level_dy}
\end{figure}

In order to understand the periodic step-like motion of the DW 
from a microscopic point of view, 
we will study the dynamics of electron state vector,
$\left|\Psi(t)\right\rangle$.
Before studying the dynamics, let us
look at the one-particle energy levels 
at $t=0$, i.e., when there is a stationary DW.  
Figure \ref{level_st} shows the
energy levels $\{E_i(0)\}$ of the Hamiltonian $\mathcal{H}(t)$
(eq. (\ref{Hamiltonian})) at $t=0$ for various values of  $J_\mathrm{sd}$.
At $J_\mathrm{sd}=0$, the energy levels are simply those of
one-dimensional tight-binding chain with a band width, $\Omega = 4$.
Each level is at least four-fold degenerate in terms of the reversal of
momentum and spin. With increasing $J_\mathrm{sd}$, the spin degeneracy
is lifted, and the energy levels gradually split into two bands. 
The lower (higher) band, 
$-(+) J_\mathrm{sd}-\Omega/2<E_i(0)<-(+)
J_\mathrm{sd}+\Omega/2$,  
is composed of the states where the spins of conduction electrons
are aligned parallel (antiparallel) to the localized spins. 
For not too large $J_\mathrm{sd}$ 
($J_\mathrm{sd}\lesssim \Omega/2$), the two bands
 have an overlapping region, 
$|E_i(0)|<-J_\mathrm{sd}+\Omega/2$.
Within each band, levels are labeled with different momentum except for
the overlapping region, 
where neither spin nor momentum serve as a good quantum number, 
i.e., the eigenstates are composed of entangled spin
and momentum degrees of freedom. This ``entangled'' energy range
will play an important role in the dynamics in the larger $K_\perp$
region, as we will show below.

To characterize the dynamics of $\left|\Psi(t)\right\rangle$, we
diagonalize the Hamiltonian $\mathcal{H}(t)$
(eq. (\ref{Hamiltonian})) at each $t$, and obtain the eigenvalues 
$\{E_i(t)\}$ and eigenvectors $\{\left|\xi_i(t)\right\rangle\}$.
We then plot the time evolution of one-particle energy
levels $\{E_i(t)\}$ and distribution functions
$\{|\left\langle\xi_i(t)|\Psi(t)\right\rangle|^2\}$ 
in Fig. \ref{level_dy}, 
corresponding to the DW motions for $K_\perp=1$ and $K_\perp =0.01$.  
First, let us discuss the case of large $K_{\perp}$ (Fig. \ref{level_dy} (a)). 
In general, the time dependence of $\{E_i(t)\}$ comes from the AB flux, 
$\Phi(t)$, 
and the localized spins, $\boldsymbol{S}_i(t)$ in eq. (\ref{Hamiltonian}). 
For $K_\perp=1$, the dynamics of $\{E_i(t)\}$ is 
mostly determined by $\Phi(t)$, since 
$\boldsymbol{S}_i(t)$ is constant except
for the moments at which $\dot{X}(t)$ shows peaks (Fig. \ref{t-v-F}).
$\Phi(t)$ accelerates the electron momentum with an acceleration rate: 
$dk/dt \sim V/N$.
As to the distribution functions, only the states below the Fermi energy
($\varepsilon_{\mathrm{F}0}\sim -1.98$) are occupied at $t=0$. 
Therefore, for small $t$, all the occupied states are within the range $E_i(t)
<J_\mathrm{sd}-\Omega/2$. In this energy range,
each state continues to be accelerated without scattering, 
since momentum
serves as a good quantum number \cite{momentum}.
On the other hand, when $E_i(t)$ 
enter the ``entangled'' energy range, $|E_i(t)|<-J_\mathrm{sd}+\Omega/2$,  
an elastic backscattering with spin flip occurs, 
and as a result, the DW acquires a recoil momentum\cite{energygap}. 
The backscattering causes rapid oscillation of 
the localized spins (see the oscillating behavior of $\dot{X}(t)$ just after the peaks in Fig. \ref{t-v-F}),  
which leads to the dissipation 
(arrow in Fig. \ref{level_dy}(a)).\cite{inelastic} 

With the difference of momentum between the successive
energy levels at $t=0$, $\delta k\sim 2\pi/N$, the interval
between the backscatterings $\delta t$ is estimated as $\delta
t\sim \delta k/(dk/dt)\sim 2\pi/V$. 
Using $V=\pi\times 10^{-4}$, we have $\delta t=2\times 10^4$, consistent
with Fig. \ref{level_dy}(a), where backscatterings  
happen at time $t_\mathrm{eb}\sim 5\times 10^4, 7\times 10^4, 9\times 10^4\cdots$.
Remarkably, the backscatterings occur exactly at the moments
at which $F(t)$ has peaks (Fig. \ref{t-v-F}). 
This correspondence shows that the backscatterings 
in the ``entangled'' energy range is the origin of 
the periodic step-like DW motion. 
Actually, the most part of DW shift $\varDelta X_\mathrm{ave}\sim 65$ (Fig. \ref{X-p}(a)), 
which is a time-averaged value of $X$ 
between $t_\mathrm{eb}-\delta t/2< t <t_\mathrm{eb}+\delta t/2$, 
can be explained by the ``momentum transfer'' at the elastic backscattering, 
$\varDelta X_\mathrm{eb}=\frac{W(c+\alpha^2)}{2\alpha(1+\alpha^2)}\varDelta k\sim 62$, 
where $\varDelta k \sim 10\delta k/2$ is a momentum shift per DW at $t_\mathrm{eb}$ 
estimated in Fig. \ref{level_dy}(a). 
It is tempting to associate the
step-like DW motion with the stick-slip motion \cite{slide}:  
``stick'' caused by the pinning due to the hard-axis anisotropy
$K_{\perp}$, and ``slip'' caused by backscatterings of the
electrons. The stick-slip motion vanishes, if the lifetime of conduction
electrons is shorter than $2\pi/V$. However, it
may be observed in a clean mesoscopic sample.

Next, we discuss the case of small $K_\perp$, i.e., 
$K_\perp=0.01$ (Fig. \ref{level_dy}(b)). 
The qualitative feature of level dynamics is distinct from that for
$K_\perp=1$. 
As soon as the electric field is applied at $t=0$, 
the energy levels of the lower band 
start to reconstruct themselves so that each level has
almost time-independent energy, except for small oscillations.
As a result, the system falls into a nonequilibrium steady state, 
where the inter-level transition rarely happens.
The distribution function of the system can be well approximated by a
shifted Fermi distribution function, with the total momentum proportional
to the applied electric field\cite{tatara1,xiao}.

In summary, we have studied the current-induced dynamics of a DW 
by solving the time-dependent Schr\"{o}dinger equation and the
LLG equations simultaneously for a one-dimensional
tight-binding chain coupled to a localized spin system. 
We observe two types of DW motions 
depending on the hard-axis anisotropy of the
localized spin system, $K_{\perp}$. For smaller $K_{\perp}$, 
the DW shows a streaming motion driven by the spin transfer. 
In contrast, for larger $K_{\perp}$, we obtain a stick-slip motion
driven by the momentum transfer. 
To our knowledge, our work provides the first
numerical realization of the crossover between the two regimes.
We have also interpreted the qualitative features of the DW
motion in both regimes, 
in terms of the dynamics of one-particle energy levels and
distribution functions. In particular, we found that the stick-slip
motion at larger $K_{\perp}$ comes from the backscatterings 
in the ``entangled'' energy range. 
Our results indicate that the DW motion at larger $K_{\perp}$ is 
influenced by the microscopic energy level structure, 
which has been neglected in most of the previous works.
The competition between the stick-slip motion 
and an extrinsic pinning or impurities 
is an interesting future problem.\cite{yamanouchi2, saitoh, hayashi, Feigenson}

The authors thank I. Maruyama and S. Kasai for fruitful discussions. 
This work was supported by the Next Generation Super Computing Project,
Nanoscience Program, MEXT, Japan. 
and also supported by GCOE for Phys. Sci. Frontier, MEXT, Japan.


\begin{thebibliography}{99}
\bibitem{torqueeffect}\ $\boldsymbol{S}_i(t)$ tends to relax parallel to
	$-\boldsymbol{H}_{\mathrm{eff}i}$ due to the Gilbert damping.
	Since the electron spin term
	$\propto -\left\langle \boldsymbol{\sigma_i} \right\rangle$
	is involved in $\boldsymbol{H}_{\mathrm{eff}i}$
	(eq. (\ref{H_eff})), $\boldsymbol{S}_i(t)$ is attracted parallel
	to $\left\langle \boldsymbol{\sigma_i}
	\right\rangle$. This is why $\tau_z(t)$ increases $\phi_0(t)$.
\bibitem{factor_c} The numerical factor $c$ is defined by the relation:\
 $\int_0^t F(t')\exp[-(t-t')/\tau_\alpha]dt'\sim cF(t)\tau_\alpha$, 
where the relaxation time due to the Gilbert damping,  
$\tau_\alpha = (1+\alpha^2)/(\alpha K_\perp)$.
\bibitem{momentum} While the translational symmetry is broken due to DW, its 
	effect is negligible except for the ``entangled'' region, then momentum 
	can be regarded as a good quantum number.
\bibitem{energygap} Even in the entangled region, the backscattering
	does not occur at every ``crossing'' of energy levels. For some
	crossings, the degree of level repulsion is small.
\bibitem{inelastic} The energy of 
localized spin system varies as $t$ with 
$\dot{E_\mathrm{S}}=
J_\mathrm{sd}\sum_{i}\dot{\boldsymbol{S}_i}\cdot
\left\langle \boldsymbol{\sigma_i} \right\rangle
-\alpha \sum_{i}\dot{\boldsymbol{S}_i}^2$. 
Firstly, the rapid oscillation of the localized spins (large
$\dot{\boldsymbol{S}_i}$) transfers the energy of electron
system to the localized spin system, through the first term of
this equation. After that, The energy of localized spins is dissipated
	due to the second term (Gilbert damping), then the state of the total system is restored to the state before the backscattering.
\end{thebibliography}
\end{document}